\documentclass[prl,twocolumn,showpacs,aps,floatfix,notitlepage]{revtex4-1}

\usepackage{latexsym,amsmath,amssymb}
\usepackage{slashed}
\usepackage{bm}
\usepackage{epsfig}
\usepackage{graphicx}

\newcommand{\be}{\begin{eqnarray}}
\newcommand{\ee}{\end{eqnarray}}
\newcommand{\rar}{\rightarrow}

\def\dd{\text{d} }

\def\+{\dagger}
\def\<{\langle}
\def\>{\rangle}

\newcommand{\cH}{{\cal H}}
\newcommand{\cA}{{\cal A}}

\newcommand{\cB}{{\cal B}}

\newcommand{\ba}{\begin{eqnarray}}
\newcommand{\ea}{\end{eqnarray}}

\newcommand{\lp}{\left(}
\newcommand{\rp}{\right)}
\newcommand{\al}{\alpha}
\newcommand{\bt}{\beta}

\newcommand{\la}{\lambda}

\newcommand{\lag}{\mathcal{L}}
\newcommand{\B}{\tilde{B}}

\begin{document}

\title{Three-magnetic fields.}

\author{Tomi~S.~Koivisto$^a$ and Federico~R.~Urban$^b$}

\affiliation{$^a$Institute for Theoretical Astrophysics, University of Oslo, P.O.\ Box 1029 Blindern, N-0315 Oslo, Norway\\
$^b$Department of Physics \& Astronomy, University of British Columbia, 6224 Agricultural Road, Vancouver, B.C. V6T 1Z1, Canada}

\date{\today}

\begin{abstract}
A completely new mechanism to generate the observed amount of large-scale cosmological magnetic fields is introduced in the context of three-form inflation.  The amplification of the fields occurs via fourth order dynamics of the vector perturbations and avoids the backreaction problem that plagues most previously introduced mechanisms. 
\end{abstract}

\pacs{98.80.Cq}

\maketitle

%%%%%%%%%%%%%%%%%%%%%%%%%%%%%%%%%%%%%%%%%%%%%%%%%%%%%%%%%%%%%%%%%%%%%%%%%%%%%%%%%%%%%%
\section{Introduction and conclusion}\label{I&C}
%%%%%%%%%%%%%%%%%%%%%%%%%%%%%%%%%%%%%%%%%%%%%%%%%%%%%%%%%%%%%%%%%%%%%%%%%%%%%%%%%%%%%%

Large scale magnetic fields exist in most observed structures in our Universe such as galaxies, 
clusters, filaments, and beyond, stretching to regions in the sky where little to next to no gas is present.  Observational evidence is growing by the day, unveiling an unsettling uniform and ubiquitous presence; yet, their nature and origin remains draped by a hazy veil of mystery, see the reviews~\cite{Han:2002ns,Subramanian:2008tt,Kandus:2010nw,Widrow:2011hs}, and~\cite{Govoni:2004as,Beck:2008ty,Neronov:1900zz,Tavecchio:2010mk,Dolag:2010ni}.

The property which is perhaps most intricate to frame within a consistent theoretical model for cosmic magnetic fields is their pervasively wide coherence length, which can stretch well beyond the Mpc mark; cosmological inflation might be the only viable candidate for its automatic transformation of short wavelength fluctuations into beyond-the-horizon correlations~\cite{Martin:2007ue}.

%Inflationary-based mechanisms however tend to be quite unsuccessful in terms of giving the large scale magnetic fields an appreciable intensity.  This is somewhat expected because, contemporaneously to exponentially stretching waves, inflation also redshifts their amplitudes, exponentially.  If the coupling between the two is engineered in such a way to overcome this damping process, then a sizeable magnetic field can be born.

Many simple options for inflationary magnetogenesis suffer from a backreaction issue, see~\cite{Demozzi:2009fu,Kanno:2009ei,Urban:2011bu,Byrnes:2011aa}.  In short, as inflation proceeds at amplifying electromagnetic waves, the overall energy density of the latter catches up with that of the former, leading to instability in the de Sitter background.  When this occurs inflation stops, and if this happens too early then we are not able to generate an almost scale invariant spectrum of curvature perturbations, as well as expand the causal region of the Universe beyond the observed one today -- a possible viable model is perhaps the simplest~\cite{Martin:2007ue}, but it appears to be successful only in the strong coupling (hence non perturbatively trustable) regime~\cite{Demozzi:2009fu,Kanno:2009ei} (but see the proposal of~\cite{Caldwell:2011ra} and criticism in~\cite{Barnaby:2012tk}).

The reasons behind this unhappy ending are twofold.  In the first place, what is generally found is that the electric field, while strongly suppressed in a highly conductive plasma (such as the primordial soup during and after reheating), plays the part of the giant in vacuum, see~\cite{Urban:2011bu}.  Secondly, the electric field tends to be redder than the magnetic field~\cite{Urban:2011bu}, and since inflation generally blindly processes an extremely wide range of modes, the magnetic spectrum needs to be close to being red or all the allowed energy density will be snatched by small scales; but if this is so, the electric field will greatly dominate in the IR, leaving only crumbs of power to the magnetic field.

In this work we show how these general issues can be circumvented in a particular model of inflation, three-form driven inflation~\cite{Germani:2009iq,Koivisto:2009sd}, thanks to the peculiar re-shaping of the magnetic spectrum the coupling between the three-form and electromagnetism produces.

Three-form inflation is a generally viable alternative to scalar field models.  Details of the rich dynamics that three-form inflation allows can be found in various cases in the references~\cite{Germani:2009iq,Koivisto:2009sd,Koivisto:2009ew,Boehmer:2011tp,Ngampitipan:2011se,Koivisto:2009fb}; for instance, the simplest models cannot produce phantom inflation without instabilities or blue-tilted spectrum, but there are viable models generating slight red tilt with or without slow roll.

Once coupled to EM, three-form inflation provides us with the two key ingredients for a promising magnetisation of the large scale Universe.  First, it entails the possibility of exponential growth in the EM sector, driven by an instability in the three-form rotational sector.  Moreover, and more importantly, provides an effective UV cutoff at which the instability dies off; it then automatically splits it from the scale of inflation, allowing for a careful selection of the modes that are amplified, thereby upsetting the democraticity of inflation.

This can be understood in the dual description of three-form inflation as a scalar field inflation, since the duality maps kinetic into potential and vice versa.  However, the duality can break down in the presence of nontrivial self-interactions~\cite{Koivisto:2009fb} of the three-form or nonminimal couplings to geometry~\cite{Germani:2009iq} or to matter~\cite{Ngampitipan:2011se}.  Thus three-form inflation is equivalently described by a scalar field model only in the simplest minimally coupled cases.  The novel aspect here is the coupling to the electromagnetic field: this both promotes the model to a physically distinct alternative to scalar-field driven constructions by breaking the duality (as it is apparent when one sees that the three-form has propagating vector degrees of freeedom), and resolves a generic shortcoming of theirs by enabling sufficient magnetic field production without the backreaction issue.

%The electric field is still the one to watch out for, as it still dominates the energy density budget.  However, its otherwise malevolous effect on the magnetic field is more than controlled by the UV cutoff.  Unitarity is also under control, for small coupling here translates in efficient amplification, unlike, for instance, the resonant model of~\cite{Byrnes:2011aa}.

Our result are shown in Figs.~\ref{fig2} and~\ref{fig2s}: despite the electric field still being the one to watch out for, magnetic fields which today have strengths in the $10^{-15}$ Gauss range can be easily produced without excessive fine-tuning of the parameters, and without jeopardising the development and completion of a long epoch of inflation.

%%%%%%%%%%%%%%%%%%%%%%%%%%%%%%%%%%%%%%%%%%%%%%%%%%%%%%%%%%%%%%%%%%%%%%%%%%%%%%%%%%%%%%
\section{The coupling}\label{threeEM}
%%%%%%%%%%%%%%%%%%%%%%%%%%%%%%%%%%%%%%%%%%%%%%%%%%%%%%%%%%%%%%%%%%%%%%%%%%%%%%%%%%%%%%

Let $A^\mu$ be the photon vector potential and $B^{\mu\nu\rho}$ the three-form. The canonical lagrangian including the both fields is
\be \label{bare}
\lag_A+\lag_B = -\frac{1}{4}F^2(A)-\frac{1}{48}F^2(B)-V(B^2)\,,
\ee
where the Faraday forms are computed from an $n$-form potential $N$ as $F(N)_{\mu_1\dots\mu_{n+1}}=(n+1)!\partial_{[\mu_1}N_{\mu_2\dots\mu_{n+1}]}$.  The components of the dual of the three-form are~\cite{Nakahara:2003nw}
\be
\B_\alpha \equiv \frac{1}{6}\epsilon_{\alpha\beta\gamma\delta}B^{\beta\gamma\delta}\,.
\ee
The most general Lorentz-invariant, quadratic, second order and $U(1)$ invariant coupling of the two fields 
is of the form
\be \label{inte}
\lag_{AB} = -\frac{1}{2}\alpha F_{\mu\nu}(A)F^{\mu\nu}(\B)\,.
%+\alpha_2(\nabla_\mu A^\mu)(\nabla_\nu \B^\nu) + (\alpha_3+\alpha_4 R)A_\mu\B^\mu
%+ \alpha_5 A^\mu R_{\mu\nu}\B^\nu + \alpha_6 A^\mu R_{\mu\nu\rho\sigma}B^{\nu\rho\sigma} \, .
\ee
\section{The amplification mechanism}\label{amplify}
%%%%%%%%%%%%%%%%%%%%%%%%%%%%%%%%%%%%%%%%%%%%%%%%%%%%%%%%%%%%%%%%%%%%%%%%%%%%%%%%%%%%%%

We decompose the spatial part of the vector potential $A_\mu=(A_0,A_i)$ in terms of its transverse and longitudinal components as $A_i=A^T_i+\partial_i\chi$, where
$\nabla\cdot {\bf A}^T = 0$.  The three-form is decomposed in a similar way where $\B_i=B^T_i+\partial_i\xi$.  Since the physical photons correspond to the transverse degrees of freedom, we are particularly interested in the vector perturbations. The vector part of the Einstein-Hilbert Lagrangian coupled to the three-form is~\cite{next}
\be\label{action0}
\lag_{EH+B}^{(v)} = \frac12 \left[ \frac{M_\text{P}^2 k^2}{2} {\bf C}^{T2} - \frac{V_{,X}}{X} \lp {\bf \B}^T - X{\bf C}^T \rp^2 \right] \, ,
\ee
where we have used the background Friedmann equation and ${\bf C}^T$ is the gauge-invariant rotational degree of freedom of the metric perturbations~\footnote{Explicitly, ${\bf C}^T={\bf c}+{\bf d}'$, when the line element is parametrised as $ds^2=a^2(\eta)[-d\eta^2+ c_i dx^i d\eta + d_{i,j}dx^i dx^j]$.};
 %\be
 %ds^2=a^2{\eta}[-d\eta^2+ {\bf c}\cdot d{\bf x}d\eta + (d{\bf x}\cdot\nabla {\bf d})\cdot d{\bf x}]\,.
 %\ee 
$X$ is the background value of the three-form field $X=\sqrt{-B^2}$ in the notation of Ref.~\cite{Koivisto:2009fb}.  Eliminating the unphysical $A_0$ through $A_0=\chi' - \al (\B_0 - \xi')$~\footnote{This is true only as long as we are interested in the $U(1)$ gauge theory; if we were to generalise~(\ref{inte}) including gauge non-invariant terms, the statement would need not apply any longer.}, the relevant part of the action~(\ref{action0}) simplifies to
%\be
%S_A + S_{AB} = \frac12 \int \left[ {\bf A}^T \lp - \partial^2_\eta + \Delta \rp \lp {\bf A}^T + 2\al {\bf \B}^T \rp - A_0 \Delta \lp A_0 - 2\chi' + 2\al \lp \B_0 - \xi' \rp \rp - \chi' \Delta \lp \chi' - 2\al \lp \B'_0 - \xi' \rp \rp \right] \dd^4 x \, .
%\ee
%Varying with respect to $A_0$, we get $A_0=\chi'-\al(\B_0-\xi')$, and the action simplifies to
\ba\label{action}
\lag_{A+AB} &=& \frac12 \Big[ {\bf A}^T \lp - \partial^2_\eta + \Delta \rp \lp {\bf A}^T + 2\al {\bf \B}^T \rp \nonumber \\
&& + \al^2 \lp \B_0 - \xi' \rp \Delta \lp \B_0 - \xi' \rp \Big] \, ,
\ea
%where we have redefined $\al_1\equiv\al$. 
where $\eta$ is conformal time, and $\Delta\equiv\delta^{ij}\partial_i\partial_j$.  The scalar polarisations of the photon are nondynamical, as expected.  It is interesting to note though that the scalar part of the interaction decouples from the electromagnetic field, it contributes to the effective sound speed of the three-form.

The equations of motion for the vector degrees of freedom become (omitting from now on the superscripts $T$, since all vectors are considered to be transverse)
\ba
M_\text{P}^2 k^2 {\bf C} + 2V_{,X}({\bf \B}-X{\bf C}) & = & 0 \, , \label{eom1} \\
V_{,X} ({\bf \B}-X{\bf C}) - \al X(-\partial_\eta^2+\Delta){\bf A} & = & 0 \, , \label{eom2} \\
(-\partial_\eta^2+\Delta)({\bf A} + 2\al{\bf \B}) & = & 0 \, . \label{eom_a}
\ea
We note that in the absence of the coupling, both the metric and three-form vector perturbations are nondynamical.  We can eliminate ${\bf C}$ and obtain a closed equation for the three-form perturbation in Fourier space:
\be\label{eom_b2}
\left[\partial_\eta^2 + \lp 1 + \frac{1}{f k^2} \rp k^2 \right] {\tilde\cB} = 0 \, ,
\ee
where
\be \label{f0}
f = 2\al^2 \frac{X}{V_{,X}} \lp \frac{2V_{,X}X}{M_\text{P}^2 k^2} - 1 \rp \, ,
\ee
and $\tilde\cB$ is the Fourier transform of $\B$.  Thus the three-form rotational modes propagate with a nontrivial dispersion relation; they can therefore in principle be significant even at large scales.  In particular, there is a divergence at $f\rar0$, that is, $2V_{,X}X \rar M_\text{P}^2 k^2$.  %At this point the effective sound speed squared will change sign, implying that the vector perturbations over some range of scales will be drastically amplified.  Such range of scales exists unless we are exactly at the origin or in a potential minimum, $V_{,X}X=0$ or consider phantom inflation $V_{,X}X<0$ which is generally unstable.
Depending on the sign of the sound speed squared, which in turn is determined by the shape of the potential, the vector perturbations over some range of scales can be drastically amplified (see below).

Plugging the solution into equation~(\ref{eom_a}) then shows that the electromagnetic potential evolves as a harmonic oscillator driven by an external force, $\cA'' + k^2 \cA = {\bf F}(\tilde\cB)$, where $F(\tilde\cB) = -2\al(\partial_\eta^2+k^2) \tilde\cB$ and $\cA$ is the Fourier transform of ${\bf A}$.  One can also consider the autonomous evolution equation for ${\bf A}$,
\be \label{eom_a2}
(-\partial_\eta^2+\Delta) \left[ 1 - f (-\partial_\eta^2+\Delta) \right] {\bf A} = 0 \, .
\ee
From this it is obvious that the usual plane waves solutions always exist,
%~\footnote{Notice that fourth order equations do not necessarily imply the presence of ghost modes (e.g.~$f(R)$ fourth order gravity); our equations~(\ref{eom1}),~(\ref{eom2}), and (\ref{eom_a}) are second order.}
but now there are two additional modes.  At $f \sim 0$, where the three-form perturbation is divergent, the equation~(\ref{eom_a2}) reduces to second order.  Going to Fourier space, we have for each propagating degree of freedom
\ba \label{eom_a3}
\cA^{(4)} & + & 2\frac{f'}{f} \cA^{(3)} + \frac{1}{f} (f''-1+2k^2f) \cA'' + 2\frac{f'}{f} k^2 \cA' \nonumber \\ & + & k^2 (\frac{f''-1}{f} + k^2) \cA = 0 \, .
\ea

Notice the appearance of fourth order time derivative in our equations, which might lead to the appearance of negative norm states.  We are working here with an effective field theory which is only valid at low energies $k \ll V_0^{1/4}$, where $V_0$ sets the UV mass scale of the model, and we need to ensure that the dynamic effects are confined to such range; we will see that this is indeed the case in our specific examples.

Hence, to summarise, the effective driving force for ${\bf A}$ induced by the instability of ${\bf \B}$ provides a mechanism to boost the magnetic fields.  We are not aware of any previous amplification mechanism based on nontrivial dynamics of the primordial rotational perturbations, since most magnetogeneses in the literature can be effectively described by introducing either a mass term or a time dependent coupling constant to the photon~\cite{Demozzi:2009fu}.

%%%%%%%%%%%%%%%%%%%%%%%%%%%%%%%%%%%%%%%%%%%%%%%%%%%%%%%%%%%%%%%%%%%%%%%%%%%%%%%%%%%%%%
\section{Examples}\label{examples}
%%%%%%%%%%%%%%%%%%%%%%%%%%%%%%%%%%%%%%%%%%%%%%%%%%%%%%%%%%%%%%%%%%%%%%%%%%%%%%%%%%%%%%

For simplicity, we consider de Sitter solutions with constant comoving field $X$. There are two classes of such fixed points~\cite{Koivisto:2009fb}: class $B$, which are the critical points corresponding to $X^2=\pm\frac{2}{3}M_\text{P}^2$, and class $C$, which are the critical points corresponding to the minima of the potential.  The stability of these fixed points depends on the shape of the potential.  In both cases the evolution equation~(\ref{eom_a3}) simplifies to
\be\label{eomSim}
\cA^{(4)} - \lp\frac{1}{f_0} - 2k^2\rp \cA'' - k^2\lp \frac{1}{f_0} - k^2 \rp \cA = 0 \, ,
\ee
$f_0$ is given by Eq.~(\ref{f0}) evaluated at the fixed point.  We can readily solve this equation to obtain
\be
\cA(\eta) = \cA_1\cos{\lp k\eta \rp} + \cA_2\sin{\lp k\eta \rp} + \cA_3 e^{\Gamma k\eta} + \cA_4 e^{-\Gamma k\eta} \, ,
\ee
where $\cA_j$ are constant vectors and $\Gamma^2 + 1 \equiv 1/(f_0 k^2)$.  Besides the usual plane waves, the two interesting and new solutions are also oscillatory if $\Gamma^2 < 0$, but if $\Gamma^2 > 0$, one of them is exponentially growing and the other exponentially decaying.

%We may denote the characteristic time scale of the instability as
%\be
%\eta_I = \sqrt{\frac{f_0}{1-f_0k^2}} = \frac{1}{\Gamma k}\,.
%\ee
%Thus an exponential instability appears due to the interaction with the three-form inflaton.

As an example consider the exponential model $V=V_0\exp{\lp - \bt X^2 / M_\text{P}^2 \rp}$.  At the fixed point $B$, which is stable if $\bt$ is positive, we have
\be\label{GammaDef}
\Gamma^2 = \frac{\kappa_\Lambda^2 - \kappa^2}{\Lambda^2 \kappa_\Lambda^2 + \kappa^2} \, ,
\ee
where we have defined $\Lambda^2 \equiv 8\al^2/3 / (1 - 8\al^2/3) \simeq 8\al^2/3$ and $k_\Lambda^2 \equiv 8\bt V / (3 M_\text{P}^2 \Lambda^2) \simeq \bt V / (\al^2 M_\text{P}^2)$.  In this way it is immediate to see that the exponentially growing solution exists only for $k \leq k_\Lambda$; $k_\Lambda$ is therefore the UV scale which defines instability band.  We will repeatedly be using the notation $\kappa \equiv k/\cH_e$ which normalises $k$ to the highest mode accessible by inflation, $\cH_e$ -- we denote $\cH \equiv a'/a$, where $a$ is the scale factor of the Universe and the prime stands for conformal time derivative.

Therefore, the scale acting as \emph{de facto} UV cutoff is identified with, $k_\Lambda$; this is well below the actual UV cutoff, set by $V_0^{1/4}$, and our theory is trustable within this limit.  In the more general and complete theory, alongside terms in the Lagrangian like $F(A)F(B)$, which would make the Hamiltonian unbounded from below, also higher order, stabilising, terms such as, schematically, $(F(A)F(B))^2 / V_0$ would appear; in the full theory, if constructed appropriately, there will be no ghosts (see for instance the Galileon case) -- we believe that it goes beyond the scope of the present work to show explicitly that such a theory can be assembled, and we shall be content with working in the, safe, low energy limit.

The other fixed point $C$ at the origin is an attractor when $\bt$ is negative. There we have
\be
\Gamma^2 k^2 = \frac{\bt V_0}{\al^2 M_\text{P}^2} - k^2 \, .
\ee
Thus, when this point is an attractor, there is no instability of the vector modes. If we are inflating at a saddle point (i.e., $\bt>0$) vector modes will be unstable at scales
$k^2<\frac{\bt V_0}{\al^2 M_\text{P}^2}$.

We summarise the properties of the fixed points for several classes of potentials and the corresponding expressions for $\Gamma^2 +1$ in table~\ref{tb2}.

\begin{table*}
\begin{tabular}{|c||c|c||c|c|}
\hline
$V(X)/V_0$ & $B$: stability & $B$: $\Gamma^2 + 1$ & $C$: stability & $C$: $\Gamma^2 + 1$  \\ \hline
$\exp(- \bt X/M_\text{P})$ & $B_{+}$ S for $\bt > 0$, $B_{-}$ S for $\bt < 0$ &$\frac{\la V_0}{2\al^2 \lp \frac43 \la V_0 \pm \sqrt{2/3} M_\text{P}^2 e^{\pm \la \sqrt{2/3}} \rp}$ 
& S & $\frac{\la V_0}{2\al^2 X M_\text{P} k^2}$ \\ \hline
$\exp(- \bt X^2 / M_\text{P}^2)$ & S for $\bt > 0$ & $\frac{\bt V_0}{\al^2 \lp \frac83 \bt V_0 + e^{-\frac23 \bt} M_\text{P}^2 k^2 \rp}$ & S for $\bt < 0$ & $ \frac{\bt V_0}{\al^2 M_\text{P}^2 k^2}$ \\ \hline
$(X/M_\text{P})^2$ & U & $\frac{V_0}{\al^2 \lp \frac83 V_0 - M_\text{P}^2 k^2 \rp}$ & S & $-\frac{V_0}{\al^2 M_\text{P}^2 k^2}$\\ \hline
$(X/M_\text{P})^{2n} , \, n>1$ & U &$\frac{n V_0}{\al^2 \lp \frac83 n V_0 - \lp \frac32 \rp^{n-1} M_\text{P}^2 k^2 \rp} $ & S & $0$\\ \hline
$\lp X^2 - C^2 \rp^2 / M_\text{P}^4$ & S for $C > \sqrt{2/3} M_\text{P}$ & $\frac{2 V_0 \lp \frac23 - C^2/M_\text{P}^2 \rp}{\al^2 \lp \frac{16}{3} V_0 \lp \frac23 - C^2/M_\text{P}^2 \rp - M_\text{P}^2 k^2 \rp}$ & $C_1$ S, $C_2$ U & $\frac{2 V_0 C^2/M_\text{P}^2}{\al^2 M_\text{P}^2 k^2}$\\ \hline
\end{tabular}
\caption{\label{tb2} Stability of the de Sitter fixed points and the characteristic width of instability of the vector potential in four classes of models. U -- unstable; S -- stable.}
\end{table*}

In general, it is easy to see when the instability for the vector potential $\cA$, parametrised by $\Gamma^2$, is present, and at which scales.  If large scales, $k\rar0$, are to be exponentially amplified, then it is necessary that $4\al^2 X^2/M_\text{P}^2 \leq 1$.  Notice that at large momenta $\Gamma^2\rar-1$ and we recover plane wave solutions.  Hence, there is in general a low energy band which is unstable given the above condition.

Beside the turning point where $\Gamma^2$ changes sign, typically there would be a singularity in $\Gamma^2$ for the mode $k$ which nullifies its denominator.  This can lie within the exponentially growing band resulting in an infinite energy density around such scale.  The question of where the singularity will occur for a specific potential depends on the sign of $V_{,X}X$: if this is positive the divergence will appear; in the opposite case $V_{,X}X < 0$ the problematic scale is beyond the UV cutoff.  In the specific examples we discuss this uncontrolled instability is avoided, whereas the exponential, controlled, growth is attained for a low energy band of modes.

%We choose then to work with the exponential potential $V = V_0 \exp(-\bt X^2/M_\text{P}^2)$ because it does not require a very high degree of fine-tuning; similar phenomenology arises in the Ginzburg-Landau case $V(X)=V_0(X^2-C^2)^2/M_\text{P}^2$, but only at the fixed point $B$.

%%%%%%%%%%%%%%%%%%%%%%%%%%%%%%%%%%%%%%%%%%%%%%%%%%%%%%%%%%%%%%%%%%%%%%%%%%%%%%%%%%%%%%
\section{The spectrum}\label{spectrum}
%%%%%%%%%%%%%%%%%%%%%%%%%%%%%%%%%%%%%%%%%%%%%%%%%%%%%%%%%%%%%%%%%%%%%%%%%%%%%%%%%%%%%%

The action~(\ref{action}) is canonical for the EM field, and as long as we treat the coupling to the three form as a perturbation we can quantise canonically~\cite{Barrow:2006ch}.  We are interested of course in solutions which grow with conformal time, so we will choose $\cA_1 = \cA_2 = \cA_4 = 0$ -- although this may appear artificial, it is immediate to convince oneself that, were these solutions kept, they would become very rapidly irrelevant in comparison to the growing $\cA_3$ term ($\cA_1$ and $\cA_2$ are simply vacuum plane waves, and $\cA_4$ decays exponentially).

We have two options in fixing the initial conditions (IC) for the EM vector potential $\cA$ in Fourier space: either we set them at some initial time, which can be taken to correspond to the beginning of inflation $\eta_i$, for all modes, or we can normalise the fields to their Bunch-Davies vacuum $\sqrt{2k} \cA = e^{-ik\eta}$ at the time they become superhorizon.  The first possibility is the simplest, but perhaps not the most physical nor appealing one, for it introduces an explicit and very strong dependence on the IC of inflation, which is antithetic to what inflation is brought in for.  Also, it poses the theoretical question on how to properly choose a vacuum state since the effective coupling is never turned off.  Nonetheless, the latter issue can be resolved if we work in the small coupling regime: in this case it is consistent to choose the Bunch-Davies solution as the approximate IC.

The other choice corresponds to a coupling which dynamically becomes relevant at some time (e.g.~horizon crossing), while being inefficient for smaller scales.  This option makes it clear that for all modes the specific IC do not matter, while they can be consistently chosen to be in vacuum at any early time without affecting the final result.  However, this demands a more elaborate theoretical motivation for the time-dependence in $\alpha$ -- if possible at all -- and will be discussed in a subsequent publication~\cite{next}.

One more comment is in order here.  Since we work with the simplified and idealised toy model where $X$ does not go anywhere, also $f$ is constant, and the equation of motion for $\tilde\cB$~(\ref{eom_b2}) does not admit plane wave solutions at any time.  However, it is easy to convince ourselves that in more realistic models, when $fk^2\rar\pm\infty$ as $k\eta\rar-\infty$ (which depends on the shape of the potential) then the rotational perturbation $\tilde\cB$ is initially in its Bunch-Davies vacuum, and so will $\cA$.  That is, the exponent $\Gamma$ in reality is time-dependent as well, and flips from imaginary to real at a given time; we take the latter to be $\eta_i$ for simplicity.  This also means that in the full model the dependence on the initial conditions will be pronouncedly less severe.

The IC fix the value of $\cA_3$ and give us the following solution to work with:
\be
\cA &=& \frac{1}{\sqrt{2k}} e^{\Gamma k(\eta_i-\eta)} \quad \text{IC at } \eta = \eta_i \, . \label{cAdef2}
\ee
%Before moving on to the actual discussion, the key definitions we will need are the magnetic power spectrum
%\be\label{deltaB}
%\delta_B^2 \equiv \frac{k^5 |\cA|^2}{4\pi^2 a^4} \, ,
%\ee
%where $a$ is the scale factor of the Universe, and the energy density in electromagnetic field:
%\be\label{energy}
%\rho_\text{EM} = \frac{1}{4\pi^2 a^4} \int \frac{\dd k}{k} k^3 \left[ |\cA'|^2 + k^2 |\cA|^2 \right] \, ,
%\ee
%where the first term is related to the electric field, and the second one is the magnetic energy density proper.  Recall that for a de Sitter solution conformal time spans the range $[-\infty,0]$, and $\cH = -1/\eta$.
We can immediately write down the corresponding expressions for the energy density associated with the EM field at the end of inflation as
\be
\rho_\text{EM} &=& \frac{1}{4\pi^2 a_e^4} \int \frac{\dd k}{k} k^3 \left[ |\cA'|^2 + k^2 |\cA|^2 \right] \label{energy} \\
&=& \frac{\cH_e^4}{8\pi^2 a_e^4} \int\dd\kappa \kappa^3 \lp \Gamma^2 + 1 \rp e^{2\Gamma\kappa(1/\kappa_i-1)} \, , \label{energy2}
\ee
where of course $\kappa_e = 1$ by definition, and we can clearly ignore the second piece in the exponent, being much smaller than the first, positive, contribution.  Also, we work with small $\alpha$, which means $\Lambda\ll1$ and $\Gamma\gg1$; this implies that in general $E^2 \gg B^2$.

The integral~(\ref{energy2}) is not simply dealt with, and we must resort to numeric methods to solve it.  The limits of integration are $\kappa = \kappa_i = \exp(-N)$, where $N$ is the total number of e-folds inflation lasts for, and $\kappa = \kappa_\Lambda$, beyond which the solution~(\ref{cAdef2}) does not apply anymore.

\begin{figure}[ht]
\centering
\includegraphics[width=0.45\textwidth]{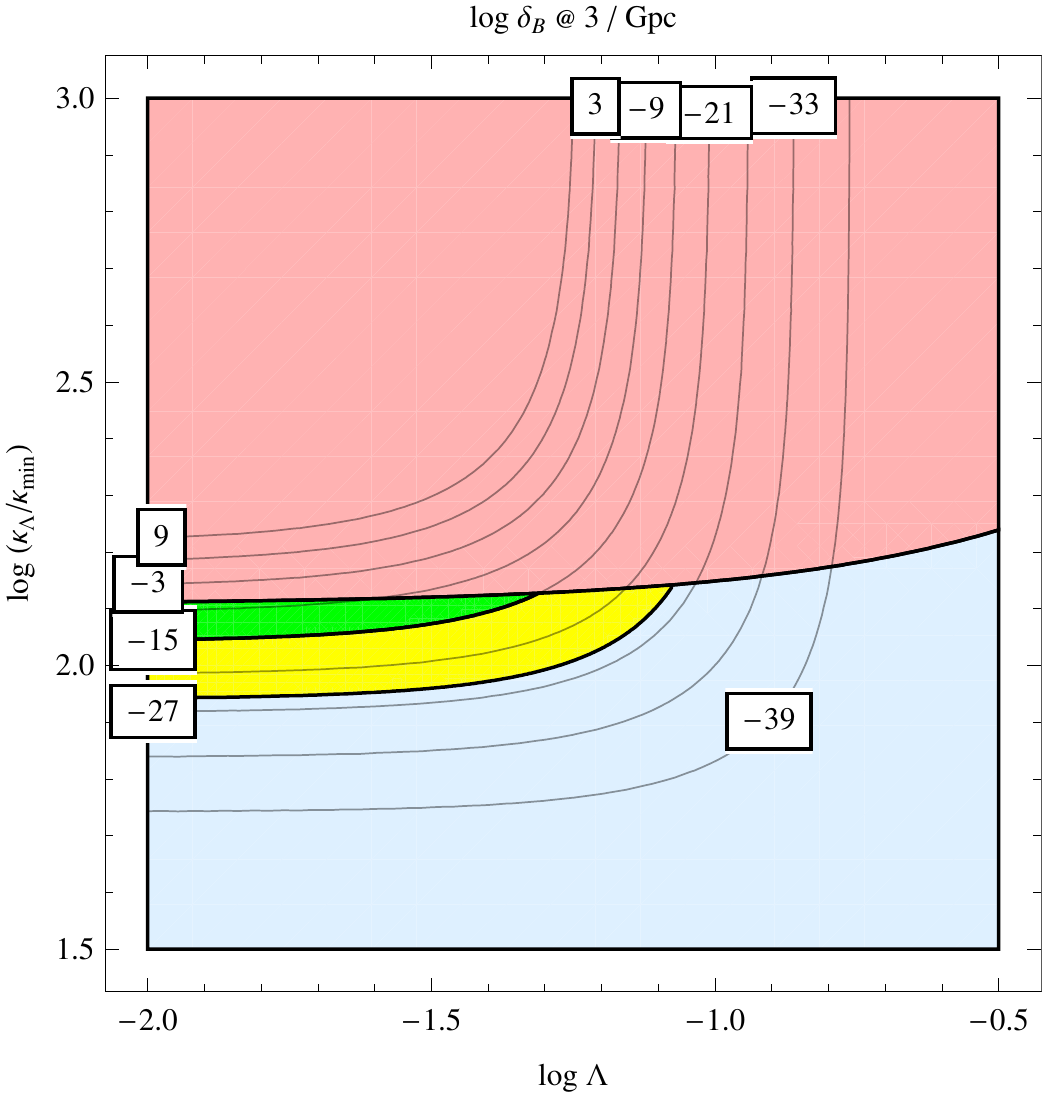}
\caption{Allowed (in light blue) and disallowed (in light red) regions in the $(\log\Lambda, \log\lp k_\Lambda/k_\text{min} \rp)$ parameters space.  Superimposed are contours of constant $\log\delta_B^0$ today, at 3/Gpc (in Gauss units).}
\label{fig2}
\end{figure}

In Fig.~\ref{fig2}, we present the regions in the parameter space $(\Lambda, k_\Lambda)$ for the well-behaved exponential potential $\exp (- \bt X^2)$, for which the ratio $\rho_\text{EM} / \rho_X$ stays below one at the end of inflation (blue region) and goes beyond one, thereby leading to instabilty in the de Sitter background (red region).  Superimposed are the contours for a given value of the magnetic power at 3/Gpc today, where the regions for which $\delta_B^0 \geq 10^{-15}$ Gauss and $\delta_B^0 \geq 10^{-25}$ Gauss are in evidence (green and yellow, respectively) -- we define $(\delta_B^0)^2 = k^5|\cA|^2 / 4\pi^2$.  These are values that correspond to the magnetic fields observed in the intergalactic medium, and a typical value for a successful seed to be fed the magnetohydrodynamic plasma at late times.

\begin{figure}[ht]
%\centering
\includegraphics[width=0.45\textwidth]{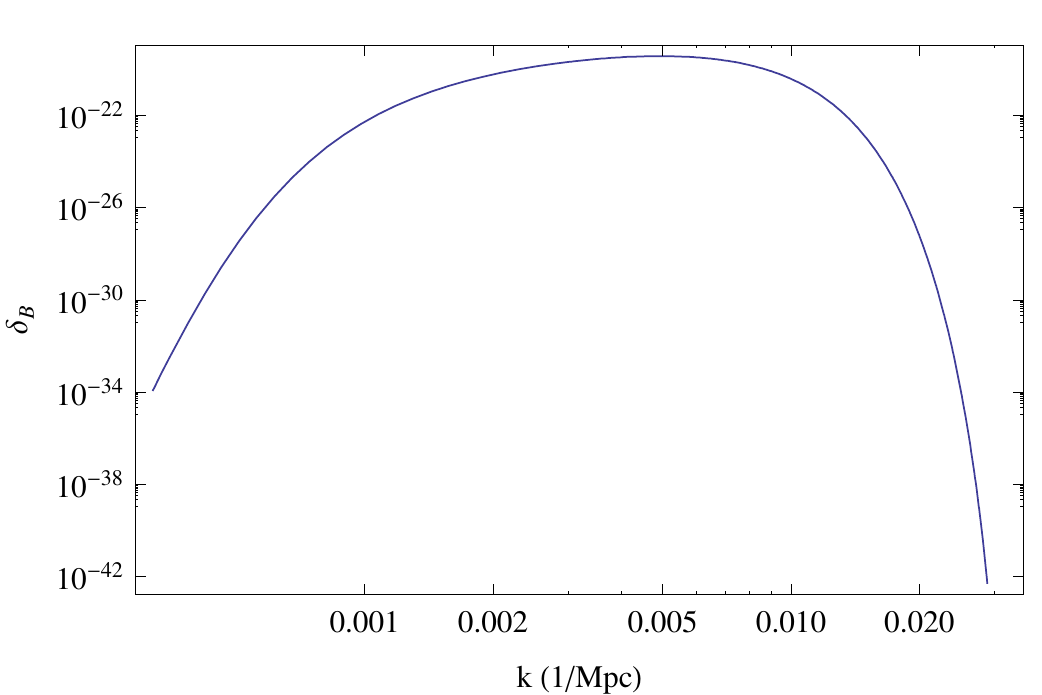}
\caption{The power spectrum today, $\delta_B^0$ (Gauss), for $\Lambda=10^{-2}$ and $k_\Lambda / k_\text{min} = 10^2$.}
\label{fig2s}
\end{figure}

Fig.~(\ref{fig2s}) plots the power spectrum today, for $\Lambda=10^{-2}$ and $k_\Lambda / k_\text{min} = 10^2$; notice the knee at high energy and the following very rapid decay.  Thanks to the UV cutoff, such a spectrum also nicely helps in avoiding the stringent constraints which late time cosmology poses on smaller scale fields~\cite{Caprini:2011cw}.  These figures were obtained for around $N=66$ e-foldings of inflation, $\cH_e/a_e \approx 10^{13}$ GeV, and assuming the Universe was always dominated by radiation from the end of inflation onwards.

\hspace{3mm}

There is some degree of fine-tuning: as one realises by looking at Eq.~(\ref{cAdef2}), these is a very strong sensitivity (doubly exponentially in terms of the total number of e-folds $N$) to the initial time, and only scales not too far from the first ones available to inflation can be amplified efficiently without leading to a catastrofic backreaction effect.  In the figure we show the final power at $k$ = 3/Gpc, where the largest mode is only around $1/\cH_0 \approx$ 4Gpc.  The plot of the spectrum makes it clear that only a couple of orders of magnitude in $k$ will be amplified efficiently; for instance, the strength at 10/Gpc is $10^{-21}$ Gauss, but drops to $10^{-47}$ Gauss at just 30/Gpc.  If we were to push the beginning of inflation further back, at much smaller $\cH_i$, it would be impossible to amplify, without backreacting, today's large scales. With the alternative prescription of a time-dependent coupling the sensitivity to initial conditions can in principle be eliminated.  Recall again also that this feature is partly due to the idealisation of the model we work with.

Looking again at the allowed region in Fig.~\ref{fig2}, it appears that one may push the coupling $\al\simeq\Lambda$ to arbitrarily small values and still come out with enough power on a given scale.  This is only partially true, since $\al$ also enters the UV cutoff, and as $\al\rar0$ one needs to keep $\al^2 k_\Lambda^2 \simeq \bt V_0 e^{-2\bt/3} / M_\text{P}^2$ fixed, that is, the window of modes which can be boosted has to be kept narrow.  Order one slopes $\bt$ would then demand extremely low bare $V_0$, which is desirable since it corresponds to little kinetic energy in the dual scalar description.  Furthermore, one can consider models with either very high or very low potential slope parameters $\bt$ to reconcile suitable UV cutoffs with small couplings $\al$. 

%%%%%%%%%%%%%%%%%%%%%%%%%%%%%%%%%%%%%%%%%%%%%%%%%%%%%%%%%%%%%%%%%%%%%%%%%%%%%%%%%%%%%%
\section*{Addendum}
%%%%%%%%%%%%%%%%%%%%%%%%%%%%%%%%%%%%%%%%%%%%%%%%%%%%%%%%%%%%%%%%%%%%%%%%%%%%%%%%%%%%%%

After this work was completed, two preprints dealing with the anisotropies generated in inflationary magnetogenesis appeared~\cite{Bonvin:2011dr,Bonvin:2011dt}.  Their conclusion is that such contributions to curvature perturbations become unacceptably large during the subsequent radiation era, thereby lapidarily coercing largest classes of models to failure.  Although we have not completed the full, and very laborious and lengthy calculation, we believe that this result needs not apply {\it as is} to our mechanism for three reasons.

The result of~\cite{Bonvin:2011dr,Bonvin:2011dt} strongly depends on the limitations on the generated power spectrum; our case presents a very peculiar spectrum, due to its low energy cutoff, which allows for almost any slope, thereby offering the possibility that the dangerous term in fact be not as big as thought.

Secondly, our Bardeen equation is quite different due to the three-form perturbations~\cite{Koivisto:2009fb}; the solutions to this equation during inflation can have different scaling with momentum $k$ and defuse the too large anisotropies.

Finally, the time evolution of $\cA$ is not a power law as in~\cite{Bonvin:2011dr,Bonvin:2011dt}, but exponential, which in turn is strongly $k$-dependent; this implies that the time-evolution of perturbations could further differ from what predicted in their case.

We elaborate on this critical issue in our upcoming publication~\cite{next}.

%{\bf [say something about the dynamics as we move away from the fixed point - is it possible to have $\bt\rar0$ or $\bt$ very large to compensate for decreasing $\al$? - viability of inflation with such parameters]}

%{\bf Conclusion.} This Letter introduced a new way to generate magnetic fields and it turned out to be viable. This also reinforces 
%the status of three-form inflation as a novel alternative to scalar fields-driven models.
%%%%%%%%%%%%%%%%%%%%%%%%%%%%%%%%%%%%%%%%%%%%%%%%%%%%%%%%%%%%%%%%%%%%%%%%%%%%%%%%%%%%%%
\section*{Acknowledgements}
%%%%%%%%%%%%%%%%%%%%%%%%%%%%%%%%%%%%%%%%%%%%%%%%%%%%%%%%%%%%%%%%%%%%%%%%%%%%%%%%%%%%%%
TK is grateful to Nelson Nunes for n-formal discussions. TK is supported by the Research Council of Norway.  FU thanks Chiara Caprini for hospitality at the IPhT, Saclay, and valuable insight.

%%%%%%%%%%%%%%%%%%%%%%%%%%%%%%%%%%%%%%%%%%%%%%%%%%%%%%%%%%%%%%%%%%%%%%%%%%%%%%%%%%%%%%
\bibliography{3fB}

\begin{thebibliography}{30}%
\makeatletter
\providecommand \@ifxundefined [1]{%
 \@ifx{#1\undefined}
}%
\providecommand \@ifnum [1]{%
 \ifnum #1\expandafter \@firstoftwo
 \else \expandafter \@secondoftwo
 \fi
}%
\providecommand \@ifx [1]{%
 \ifx #1\expandafter \@firstoftwo
 \else \expandafter \@secondoftwo
 \fi
}%
\providecommand \natexlab [1]{#1}%
\providecommand \enquote  [1]{``#1''}%
\providecommand \bibnamefont  [1]{#1}%
\providecommand \bibfnamefont [1]{#1}%
\providecommand \citenamefont [1]{#1}%
\providecommand \href@noop [0]{\@secondoftwo}%
\providecommand \href [0]{\begingroup \@sanitize@url \@href}%
\providecommand \@href[1]{\@@startlink{#1}\@@href}%
\providecommand \@@href[1]{\endgroup#1\@@endlink}%
\providecommand \@sanitize@url [0]{\catcode `\\12\catcode `\$12\catcode
  `\&12\catcode `\#12\catcode `\^12\catcode `\_12\catcode `\%12\relax}%
\providecommand \@@startlink[1]{}%
\providecommand \@@endlink[0]{}%
\providecommand \url  [0]{\begingroup\@sanitize@url \@url }%
\providecommand \@url [1]{\endgroup\@href {#1}{\urlprefix }}%
\providecommand \urlprefix  [0]{URL }%
\providecommand \Eprint [0]{\href }%
\providecommand \doibase [0]{http://dx.doi.org/}%
\providecommand \selectlanguage [0]{\@gobble}%
\providecommand \bibinfo  [0]{\@secondoftwo}%
\providecommand \bibfield  [0]{\@secondoftwo}%
\providecommand \translation [1]{[#1]}%
\providecommand \BibitemOpen [0]{}%
\providecommand \bibitemStop [0]{}%
\providecommand \bibitemNoStop [0]{.\EOS\space}%
\providecommand \EOS [0]{\spacefactor3000\relax}%
\providecommand \BibitemShut  [1]{\csname bibitem#1\endcsname}%
\let\auto@bib@innerbib\@empty
%</preamble>
\bibitem [{\citenamefont {Han}\ and\ \citenamefont
  {Wielebinski}(2002)}]{Han:2002ns}%
  \BibitemOpen
  \bibfield  {author} {\bibinfo {author} {\bibfnamefont {J.-L.}\ \bibnamefont
  {Han}}\ and\ \bibinfo {author} {\bibfnamefont {R.}~\bibnamefont
  {Wielebinski}},\ }\href {\doibase 10.1088/1009-9271/2/4/293} {\bibfield
  {journal} {\bibinfo  {journal} {Chin. J. Astron. Astrophys.}\ }\textbf
  {\bibinfo {volume} {2}},\ \bibinfo {pages} {293} (\bibinfo {year} {2002})},\
  \Eprint {http://arxiv.org/abs/astro-ph/0209090} {arXiv:astro-ph/0209090}
  \BibitemShut {NoStop}%
%%CITATION = ASTRO-PH/0209090;%%
\bibitem [{\citenamefont {Subramanian}(2007)}]{Subramanian:2008tt}%
  \BibitemOpen
  \bibfield  {author} {\bibinfo {author} {\bibfnamefont {K.}~\bibnamefont
  {Subramanian}},\ }\href@noop {} {\bibfield  {journal} {\bibinfo  {journal}
  {PoS}\ }\textbf {\bibinfo {volume} {MRU}},\ \bibinfo {pages} {071} (\bibinfo
  {year} {2007})},\ \Eprint {http://arxiv.org/abs/0802.2804} {arXiv:0802.2804
  [astro-ph]} \BibitemShut {NoStop}%
%%CITATION = 0802.2804;%%
\bibitem [{\citenamefont {Kandus}\ \emph {et~al.}(2011)\citenamefont {Kandus},
  \citenamefont {Kunze},\ and\ \citenamefont {Tsagas}}]{Kandus:2010nw}%
  \BibitemOpen
  \bibfield  {author} {\bibinfo {author} {\bibfnamefont {A.}~\bibnamefont
  {Kandus}}, \bibinfo {author} {\bibfnamefont {K.~E.}\ \bibnamefont {Kunze}}, \
  and\ \bibinfo {author} {\bibfnamefont {C.~G.}\ \bibnamefont {Tsagas}},\
  }\href {\doibase 10.1016/j.physrep.2011.03.001} {\bibfield  {journal}
  {\bibinfo  {journal} {Phys. Rept.}\ }\textbf {\bibinfo {volume} {505}},\
  \bibinfo {pages} {1} (\bibinfo {year} {2011})},\ \Eprint
  {http://arxiv.org/abs/1007.3891} {arXiv:1007.3891 [astro-ph.CO]} \BibitemShut
  {NoStop}%
%%CITATION = 1007.3891;%%
\bibitem [{\citenamefont {Widrow}\ \emph {et~al.}(2011)\citenamefont {Widrow}
  \emph {et~al.}}]{Widrow:2011hs}%
  \BibitemOpen
  \bibfield  {author} {\bibinfo {author} {\bibfnamefont {L.~M.}\ \bibnamefont
  {Widrow}} \emph {et~al.},\ }\href@noop {} {\  (\bibinfo {year} {2011})},\
  \Eprint {http://arxiv.org/abs/1109.4052} {arXiv:1109.4052 [astro-ph.CO]}
  \BibitemShut {NoStop}%
%%CITATION = 1109.4052;%%
\bibitem [{\citenamefont {Govoni}\ and\ \citenamefont
  {Feretti}(2004)}]{Govoni:2004as}%
  \BibitemOpen
  \bibfield  {author} {\bibinfo {author} {\bibfnamefont {F.}~\bibnamefont
  {Govoni}}\ and\ \bibinfo {author} {\bibfnamefont {L.}~\bibnamefont
  {Feretti}},\ }\href {\doibase 10.1142/S0218271804005080} {\bibfield
  {journal} {\bibinfo  {journal} {Int. J. Mod. Phys.}\ }\textbf {\bibinfo
  {volume} {D13}},\ \bibinfo {pages} {1549} (\bibinfo {year} {2004})},\ \Eprint
  {http://arxiv.org/abs/astro-ph/0410182} {arXiv:astro-ph/0410182} \BibitemShut
  {NoStop}%
%%CITATION = ASTRO-PH/0410182;%%
\bibitem [{\citenamefont {Beck}(2009)}]{Beck:2008ty}%
  \BibitemOpen
  \bibfield  {author} {\bibinfo {author} {\bibfnamefont {R.}~\bibnamefont
  {Beck}},\ }\href {\doibase 10.1063/1.3076806} {\bibfield  {journal} {\bibinfo
   {journal} {AIP Conf. Proc.}\ }\textbf {\bibinfo {volume} {1085}},\ \bibinfo
  {pages} {83} (\bibinfo {year} {2009})},\ \Eprint
  {http://arxiv.org/abs/0810.2923} {arXiv:0810.2923 [astro-ph]} \BibitemShut
  {NoStop}%
%%CITATION = 0810.2923;%%
\bibitem [{\citenamefont {Neronov}\ and\ \citenamefont
  {Vovk}(2010)}]{Neronov:1900zz}%
  \BibitemOpen
  \bibfield  {author} {\bibinfo {author} {\bibfnamefont {A.}~\bibnamefont
  {Neronov}}\ and\ \bibinfo {author} {\bibfnamefont {I.}~\bibnamefont {Vovk}},\
  }\href {\doibase 10.1126/science.1184192} {\bibfield  {journal} {\bibinfo
  {journal} {Science}\ }\textbf {\bibinfo {volume} {328}},\ \bibinfo {pages}
  {73} (\bibinfo {year} {2010})},\ \Eprint {http://arxiv.org/abs/1006.3504}
  {arXiv:1006.3504 [astro-ph.HE]} \BibitemShut {NoStop}%
%%CITATION = 1006.3504;%%
\bibitem [{\citenamefont {Tavecchio}\ \emph {et~al.}(2010)\citenamefont
  {Tavecchio} \emph {et~al.}}]{Tavecchio:2010mk}%
  \BibitemOpen
  \bibfield  {author} {\bibinfo {author} {\bibfnamefont {F.}~\bibnamefont
  {Tavecchio}} \emph {et~al.},\ }\href {\doibase
  10.1111/j.1745-3933.2010.00884.x} {\bibfield  {journal} {\bibinfo  {journal}
  {Mon. Not. Roy. Astron. Soc.}\ }\textbf {\bibinfo {volume} {406}},\ \bibinfo
  {pages} {L70} (\bibinfo {year} {2010})},\ \Eprint
  {http://arxiv.org/abs/1004.1329} {arXiv:1004.1329 [astro-ph.CO]} \BibitemShut
  {NoStop}%
%%CITATION = 1004.1329;%%
\bibitem [{\citenamefont {Dolag}\ \emph {et~al.}(2011)\citenamefont {Dolag},
  \citenamefont {Kachelriess}, \citenamefont {Ostapchenko},\ and\ \citenamefont
  {Tomas}}]{Dolag:2010ni}%
  \BibitemOpen
  \bibfield  {author} {\bibinfo {author} {\bibfnamefont {K.}~\bibnamefont
  {Dolag}}, \bibinfo {author} {\bibfnamefont {M.}~\bibnamefont {Kachelriess}},
  \bibinfo {author} {\bibfnamefont {S.}~\bibnamefont {Ostapchenko}}, \ and\
  \bibinfo {author} {\bibfnamefont {R.}~\bibnamefont {Tomas}},\ }\href@noop {}
  {\bibfield  {journal} {\bibinfo  {journal} {Astrophys. J.}\ }\textbf
  {\bibinfo {volume} {727}},\ \bibinfo {pages} {L4} (\bibinfo {year} {2011})},\
  \Eprint {http://arxiv.org/abs/1009.1782} {arXiv:1009.1782 [astro-ph.HE]}
  \BibitemShut {NoStop}%
%%CITATION = 1009.1782;%%
\bibitem [{\citenamefont {Martin}\ and\ \citenamefont
  {Yokoyama}(2008)}]{Martin:2007ue}%
  \BibitemOpen
  \bibfield  {author} {\bibinfo {author} {\bibfnamefont {J.}~\bibnamefont
  {Martin}}\ and\ \bibinfo {author} {\bibfnamefont {J.}~\bibnamefont
  {Yokoyama}},\ }\href {\doibase 10.1088/1475-7516/2008/01/025} {\bibfield
  {journal} {\bibinfo  {journal} {JCAP}\ }\textbf {\bibinfo {volume} {0801}},\
  \bibinfo {pages} {025} (\bibinfo {year} {2008})},\ \Eprint
  {http://arxiv.org/abs/0711.4307} {arXiv:0711.4307 [astro-ph]} \BibitemShut
  {NoStop}%
%%CITATION = 0711.4307;%%
\bibitem [{\citenamefont {Demozzi}\ \emph {et~al.}(2009)\citenamefont
  {Demozzi}, \citenamefont {Mukhanov},\ and\ \citenamefont
  {Rubinstein}}]{Demozzi:2009fu}%
  \BibitemOpen
  \bibfield  {author} {\bibinfo {author} {\bibfnamefont {V.}~\bibnamefont
  {Demozzi}}, \bibinfo {author} {\bibfnamefont {V.}~\bibnamefont {Mukhanov}}, \
  and\ \bibinfo {author} {\bibfnamefont {H.}~\bibnamefont {Rubinstein}},\
  }\href {\doibase 10.1088/1475-7516/2009/08/025} {\bibfield  {journal}
  {\bibinfo  {journal} {JCAP}\ }\textbf {\bibinfo {volume} {0908}},\ \bibinfo
  {pages} {025} (\bibinfo {year} {2009})},\ \Eprint
  {http://arxiv.org/abs/0907.1030} {arXiv:0907.1030 [astro-ph.CO]} \BibitemShut
  {NoStop}%
%%CITATION = 0907.1030;%%
\bibitem [{\citenamefont {Kanno}\ \emph {et~al.}(2009)\citenamefont {Kanno},
  \citenamefont {Soda},\ and\ \citenamefont {Watanabe}}]{Kanno:2009ei}%
  \BibitemOpen
  \bibfield  {author} {\bibinfo {author} {\bibfnamefont {S.}~\bibnamefont
  {Kanno}}, \bibinfo {author} {\bibfnamefont {J.}~\bibnamefont {Soda}}, \ and\
  \bibinfo {author} {\bibfnamefont {M.-a.}\ \bibnamefont {Watanabe}},\ }\href
  {\doibase 10.1088/1475-7516/2009/12/009} {\bibfield  {journal} {\bibinfo
  {journal} {JCAP}\ }\textbf {\bibinfo {volume} {0912}},\ \bibinfo {pages}
  {009} (\bibinfo {year} {2009})},\ \Eprint {http://arxiv.org/abs/0908.3509}
  {arXiv:0908.3509 [astro-ph.CO]} \BibitemShut {NoStop}%
%%CITATION = 0908.3509;%%
\bibitem [{\citenamefont {Urban}(2011)}]{Urban:2011bu}%
  \BibitemOpen
  \bibfield  {author} {\bibinfo {author} {\bibfnamefont {F.~R.}\ \bibnamefont
  {Urban}},\ }\href {\doibase 10.1088/1475-7516/2011/12/012} {\bibfield
  {journal} {\bibinfo  {journal} {JCAP}\ }\textbf {\bibinfo {volume} {1112}},\
  \bibinfo {pages} {012} (\bibinfo {year} {2011})},\ \Eprint
  {http://arxiv.org/abs/1111.1006} {arXiv:1111.1006 [astro-ph.CO]} \BibitemShut
  {NoStop}%
%%CITATION = ARXIV:1111.1006;%%
\bibitem [{\citenamefont {Byrnes}\ \emph {et~al.}(2012)\citenamefont {Byrnes},
  \citenamefont {Hollenstein}, \citenamefont {Jain},\ and\ \citenamefont
  {Urban}}]{Byrnes:2011aa}%
  \BibitemOpen
  \bibfield  {author} {\bibinfo {author} {\bibfnamefont {C.~T.}\ \bibnamefont
  {Byrnes}}, \bibinfo {author} {\bibfnamefont {L.}~\bibnamefont {Hollenstein}},
  \bibinfo {author} {\bibfnamefont {R.~K.}\ \bibnamefont {Jain}}, \ and\
  \bibinfo {author} {\bibfnamefont {F.~R.}\ \bibnamefont {Urban}},\ }\href
  {\doibase 10.1088/1475-7516/2012/03/009} {\bibfield  {journal} {\bibinfo
  {journal} {JCAP}\ }\textbf {\bibinfo {volume} {1203}},\ \bibinfo {pages}
  {009} (\bibinfo {year} {2012})},\ \Eprint {http://arxiv.org/abs/1111.2030}
  {arXiv:1111.2030 [astro-ph.CO]} \BibitemShut {NoStop}%
%%CITATION = ARXIV:1111.2030;%%
\bibitem [{\citenamefont {Caldwell}\ \emph {et~al.}(2011)\citenamefont
  {Caldwell}, \citenamefont {Motta},\ and\ \citenamefont
  {Kamionkowski}}]{Caldwell:2011ra}%
  \BibitemOpen
  \bibfield  {author} {\bibinfo {author} {\bibfnamefont {R.~R.}\ \bibnamefont
  {Caldwell}}, \bibinfo {author} {\bibfnamefont {L.}~\bibnamefont {Motta}}, \
  and\ \bibinfo {author} {\bibfnamefont {M.}~\bibnamefont {Kamionkowski}},\
  }\href {\doibase 10.1103/PhysRevD.84.123525} {\bibfield  {journal} {\bibinfo
  {journal} {Phys.Rev.}\ }\textbf {\bibinfo {volume} {D84}},\ \bibinfo {pages}
  {123525} (\bibinfo {year} {2011})},\ \bibinfo {note} {11 pages, 3 pdf
  figures},\ \Eprint {http://arxiv.org/abs/1109.4415} {arXiv:1109.4415
  [astro-ph.CO]} \BibitemShut {NoStop}%
%%CITATION = ARXIV:1109.4415;%%
\bibitem [{\citenamefont {Barnaby}\ \emph {et~al.}(2012)\citenamefont
  {Barnaby}, \citenamefont {Namba},\ and\ \citenamefont
  {Peloso}}]{Barnaby:2012tk}%
  \BibitemOpen
  \bibfield  {author} {\bibinfo {author} {\bibfnamefont {N.}~\bibnamefont
  {Barnaby}}, \bibinfo {author} {\bibfnamefont {R.}~\bibnamefont {Namba}}, \
  and\ \bibinfo {author} {\bibfnamefont {M.}~\bibnamefont {Peloso}},\
  }\href@noop {} {\  (\bibinfo {year} {2012})},\ \bibinfo {note} {25 pages, no
  figures. Improved discussion of the shape of the bispectrum. References
  added},\ \Eprint {http://arxiv.org/abs/1202.1469} {arXiv:1202.1469
  [astro-ph.CO]} \BibitemShut {NoStop}%
%%CITATION = ARXIV:1202.1469;%%
\bibitem [{\citenamefont {Germani}\ and\ \citenamefont
  {Kehagias}(2009)}]{Germani:2009iq}%
  \BibitemOpen
  \bibfield  {author} {\bibinfo {author} {\bibfnamefont {C.}~\bibnamefont
  {Germani}}\ and\ \bibinfo {author} {\bibfnamefont {A.}~\bibnamefont
  {Kehagias}},\ }\href {\doibase 10.1088/1475-7516/2009/03/028} {\bibfield
  {journal} {\bibinfo  {journal} {JCAP}\ }\textbf {\bibinfo {volume} {0903}},\
  \bibinfo {pages} {028} (\bibinfo {year} {2009})},\ \bibinfo {note} {* Brief
  entry *},\ \Eprint {http://arxiv.org/abs/0902.3667} {arXiv:0902.3667
  [astro-ph.CO]} \BibitemShut {NoStop}%
\bibitem [{\citenamefont {Koivisto}\ \emph {et~al.}(2009)\citenamefont
  {Koivisto}, \citenamefont {Mota},\ and\ \citenamefont
  {Pitrou}}]{Koivisto:2009sd}%
  \BibitemOpen
  \bibfield  {author} {\bibinfo {author} {\bibfnamefont {T.~S.}\ \bibnamefont
  {Koivisto}}, \bibinfo {author} {\bibfnamefont {D.~F.}\ \bibnamefont {Mota}},
  \ and\ \bibinfo {author} {\bibfnamefont {C.}~\bibnamefont {Pitrou}},\ }\href
  {\doibase 10.1088/1126-6708/2009/09/092} {\bibfield  {journal} {\bibinfo
  {journal} {JHEP}\ }\textbf {\bibinfo {volume} {0909}},\ \bibinfo {pages}
  {092} (\bibinfo {year} {2009})},\ \bibinfo {note} {* Brief entry *},\ \Eprint
  {http://arxiv.org/abs/0903.4158} {arXiv:0903.4158 [astro-ph.CO]} \BibitemShut
  {NoStop}%
\bibitem [{\citenamefont {Koivisto}\ and\ \citenamefont
  {Nunes}(2010)}]{Koivisto:2009ew}%
  \BibitemOpen
  \bibfield  {author} {\bibinfo {author} {\bibfnamefont {T.~S.}\ \bibnamefont
  {Koivisto}}\ and\ \bibinfo {author} {\bibfnamefont {N.~J.}\ \bibnamefont
  {Nunes}},\ }\href {\doibase 10.1016/j.physletb.2010.01.051} {\bibfield
  {journal} {\bibinfo  {journal} {Phys.Lett.}\ }\textbf {\bibinfo {volume}
  {B685}},\ \bibinfo {pages} {105} (\bibinfo {year} {2010})},\ \Eprint
  {http://arxiv.org/abs/0907.3883} {arXiv:0907.3883 [astro-ph.CO]} \BibitemShut
  {NoStop}%
\bibitem [{\citenamefont {Boehmer}\ \emph {et~al.}(2011)\citenamefont
  {Boehmer}, \citenamefont {Chan},\ and\ \citenamefont
  {Lazkoz}}]{Boehmer:2011tp}%
  \BibitemOpen
  \bibfield  {author} {\bibinfo {author} {\bibfnamefont {C.~G.}\ \bibnamefont
  {Boehmer}}, \bibinfo {author} {\bibfnamefont {N.}~\bibnamefont {Chan}}, \
  and\ \bibinfo {author} {\bibfnamefont {R.}~\bibnamefont {Lazkoz}},\
  }\href@noop {} {\  (\bibinfo {year} {2011})},\ \Eprint
  {http://arxiv.org/abs/1111.6247} {arXiv:1111.6247 [gr-qc]} \BibitemShut
  {NoStop}%
%%CITATION = ARXIV:1111.6247;%%
\bibitem [{\citenamefont {Ngampitipan}\ and\ \citenamefont
  {Wongjun}(2011)}]{Ngampitipan:2011se}%
  \BibitemOpen
  \bibfield  {author} {\bibinfo {author} {\bibfnamefont {T.}~\bibnamefont
  {Ngampitipan}}\ and\ \bibinfo {author} {\bibfnamefont {P.}~\bibnamefont
  {Wongjun}},\ }\href {\doibase 10.1088/1475-7516/2011/11/036} {\bibfield
  {journal} {\bibinfo  {journal} {JCAP}\ }\textbf {\bibinfo {volume} {1111}},\
  \bibinfo {pages} {036} (\bibinfo {year} {2011})},\ \Eprint
  {http://arxiv.org/abs/1108.0140} {arXiv:1108.0140 [hep-ph]} \BibitemShut
  {NoStop}%
%%CITATION = ARXIV:1108.0140;%%
\bibitem [{\citenamefont {Koivisto}\ and\ \citenamefont
  {Nunes}(2009)}]{Koivisto:2009fb}%
  \BibitemOpen
  \bibfield  {author} {\bibinfo {author} {\bibfnamefont {T.~S.}\ \bibnamefont
  {Koivisto}}\ and\ \bibinfo {author} {\bibfnamefont {N.~J.}\ \bibnamefont
  {Nunes}},\ }\href {\doibase 10.1103/PhysRevD.80.103509} {\bibfield  {journal}
  {\bibinfo  {journal} {Phys. Rev.}\ }\textbf {\bibinfo {volume} {D80}},\
  \bibinfo {pages} {103509} (\bibinfo {year} {2009})},\ \Eprint
  {http://arxiv.org/abs/0908.0920} {arXiv:0908.0920 [astro-ph.CO]} \BibitemShut
  {NoStop}%
%%CITATION = 0908.0920;%%
\bibitem [{\citenamefont {Nakahara}(2003)}]{Nakahara:2003nw}%
  \BibitemOpen
  \bibfield  {author} {\bibinfo {author} {\bibfnamefont {M.}~\bibnamefont
  {Nakahara}},\ }\href@noop {} {\  (\bibinfo {year} {2003})},\ \bibinfo {note}
  {boca Raton, USA: Taylor \& Francis (2003) 573 p}\BibitemShut {NoStop}%
\bibitem [{\citenamefont {Koivisto}\ and\ \citenamefont {Urban}(2012)}]{next}%
  \BibitemOpen
  \bibfield  {author} {\bibinfo {author} {\bibfnamefont {T.}~\bibnamefont
  {Koivisto}}\ and\ \bibinfo {author} {\bibfnamefont {F.}~\bibnamefont
  {Urban}},\ }\href@noop {} {\  (\bibinfo {year} {2012})}\BibitemShut {NoStop}%
\bibitem [{Note1()}]{Note1}%
  \BibitemOpen
  \bibinfo {note} {Explicitly, ${\protect \bf C}^T={\protect \bf c}+{\protect
  \bf d}'$, when the line element is parametrised as $ds^2=a^2(\eta )[-d\eta
  ^2+ c_i dx^i d\eta + d_{i,j}dx^i dx^j]$.}\BibitemShut {Stop}%
\bibitem [{Note2()}]{Note2}%
  \BibitemOpen
  \bibinfo {note} {This is true only as long as we are interested in the $U(1)$
  gauge theory; if we were to generalise~(\ref {inte}) including gauge
  non-invariant terms, the statement would need not apply any
  longer.}\BibitemShut {Stop}%
\bibitem [{\citenamefont {Barrow}\ \emph {et~al.}(2007)\citenamefont {Barrow},
  \citenamefont {Maartens},\ and\ \citenamefont {Tsagas}}]{Barrow:2006ch}%
  \BibitemOpen
  \bibfield  {author} {\bibinfo {author} {\bibfnamefont {J.~D.}\ \bibnamefont
  {Barrow}}, \bibinfo {author} {\bibfnamefont {R.}~\bibnamefont {Maartens}}, \
  and\ \bibinfo {author} {\bibfnamefont {C.~G.}\ \bibnamefont {Tsagas}},\
  }\href {\doibase 10.1016/j.physrep.2007.04.006} {\bibfield  {journal}
  {\bibinfo  {journal} {Phys. Rept.}\ }\textbf {\bibinfo {volume} {449}},\
  \bibinfo {pages} {131} (\bibinfo {year} {2007})},\ \Eprint
  {http://arxiv.org/abs/astro-ph/0611537} {arXiv:astro-ph/0611537} \BibitemShut
  {NoStop}%
%%CITATION = ASTRO-PH/0611537;%%
\bibitem [{\citenamefont {Caprini}(2010)}]{Caprini:2011cw}%
  \BibitemOpen
  \bibfield  {author} {\bibinfo {author} {\bibfnamefont {C.}~\bibnamefont
  {Caprini}},\ }\href@noop {} {\bibfield  {journal} {\bibinfo  {journal} {PoS}\
  }\textbf {\bibinfo {volume} {TEXAS2010}},\ \bibinfo {pages} {222} (\bibinfo
  {year} {2010})},\ \Eprint {http://arxiv.org/abs/1103.4060} {arXiv:1103.4060
  [astro-ph.CO]} \BibitemShut {NoStop}%
%%CITATION = 1103.4060;%%
\bibitem [{\citenamefont {Bonvin}\ \emph
  {et~al.}(2011{\natexlab{a}})\citenamefont {Bonvin}, \citenamefont {Caprini},\
  and\ \citenamefont {Durrer}}]{Bonvin:2011dr}%
  \BibitemOpen
  \bibfield  {author} {\bibinfo {author} {\bibfnamefont {C.}~\bibnamefont
  {Bonvin}}, \bibinfo {author} {\bibfnamefont {C.}~\bibnamefont {Caprini}}, \
  and\ \bibinfo {author} {\bibfnamefont {R.}~\bibnamefont {Durrer}},\
  }\href@noop {} {\  (\bibinfo {year} {2011}{\natexlab{a}})},\ \Eprint
  {http://arxiv.org/abs/1112.3897} {arXiv:1112.3897 [astro-ph.CO]} \BibitemShut
  {NoStop}%
%%CITATION = ARXIV:1112.3897;%%
\bibitem [{\citenamefont {Bonvin}\ \emph
  {et~al.}(2011{\natexlab{b}})\citenamefont {Bonvin}, \citenamefont {Caprini},\
  and\ \citenamefont {Durrer}}]{Bonvin:2011dt}%
  \BibitemOpen
  \bibfield  {author} {\bibinfo {author} {\bibfnamefont {C.}~\bibnamefont
  {Bonvin}}, \bibinfo {author} {\bibfnamefont {C.}~\bibnamefont {Caprini}}, \
  and\ \bibinfo {author} {\bibfnamefont {R.}~\bibnamefont {Durrer}},\
  }\href@noop {} {\  (\bibinfo {year} {2011}{\natexlab{b}})},\ \Eprint
  {http://arxiv.org/abs/1112.3901} {arXiv:1112.3901 [astro-ph.CO]} \BibitemShut
  {NoStop}%
%%CITATION = ARXIV:1112.3901;%%
\end{thebibliography}%

\end{document}